\documentclass[aps,prab,preprint,superscriptaddress]{revtex4-2}
\usepackage{graphicx}
\usepackage{algorithm2e}

\begin{document}


\title{Bayesian optimization of beam injection and storage in the PSI muEDM Experiment}



\author{Siew~Yan~Hoh}
\thanks{These authors contributed equally to this work}
\author{Yuzhi~Shang}
\thanks{These authors contributed equally to this work}
\affiliation{Tsung-Dao Lee Institute and School of Physics and Astronomy, Shanghai Jiao Tong University, Shanghai 201210, China}

\author{Ritwika Chakraborty}
\affiliation{Paul Scherrer Institut, Forschungsstrasse 111, 5232 Villigen PSI, Switzerland}

\author{Tianqi~Hu}
\affiliation{Tsung-Dao Lee Institute and School of Physics and Astronomy, Shanghai Jiao Tong University, Shanghai 201210, China}

\author{Timothy~Hume}
\affiliation{Paul Scherrer Institut, Forschungsstrasse 111, 5232 Villigen PSI, Switzerland}

\author{Kim~Siang~Khaw}
\email{kimsiang84@sjtu.edu.cn}
\affiliation{Tsung-Dao Lee Institute and School of Physics and Astronomy, Shanghai Jiao Tong University, Shanghai 201210, China}

\author{Philipp Schmidt-Wellenburg}
\affiliation{Paul Scherrer Institut, Forschungsstrasse 111, 5232 Villigen PSI, Switzerland}

\author{Yusuke~Takeuchi}
\author{Guan~Ming~Wong}
\affiliation{Tsung-Dao Lee Institute and School of Physics and Astronomy, Shanghai Jiao Tong University, Shanghai 201210, China}

\date{\today}

\begin{abstract}
The muEDM experiment at the Paul Scherrer Institute aims to measure the electric dipole moment with an unprecedented sensitivity of $6 \times 10^{-23}\,e\cdot\mathrm{cm}$. A key aspect of this experiment is the injection and storage of the muon beam, which traverses a long, narrow superconducting channel before entering a solenoid magnet. A pulsed magnetic field then kicks the muon into a stable orbit within the solenoid's central region, where the electric dipole moment is measured. To study the beam injection and storage process, we developed a G4beamline simulation to model the dynamics of beam injection and storage, incorporating all relevant electric and magnetic fields. We subsequently employed a Bayesian optimization technique to improve the muon storage efficiency for Phase I of the muEDM experiment. The optimization is demonstrated using data simulated by G4beamline. We have observed an enhancement in the beam injection and storage efficiency, which increased to 0.592\% through using Bayesian optimization with Gaussian processes, compared to 0.324\% when employing the polynomial chaos expansion. This approach can be applied to adjust experimental parameters, aiding in achieving the desired beam injection and storage performance in the muEDM experiment.
\end{abstract}

\maketitle

\section{Introduction}
In the context of accelerator and storage ring experiments, optimizing the beam injection and storage processes in both simulations and real-world applications necessitates considerable computational power and time resources. This requirement renders the optimization landscape intricate and frequently non-linear, characterized by numerous interacting parameters. Furthermore, this complexity is intensified by operational noise and the resource-demanding nature of beam physics simulations. In precision muon physics experiments, such as the Muon $g-2$ experiment~\cite{Muong-2:2023cdq, Muong-2:2024hpx} at Fermilab and the muEDM experiment~\cite{Adelmann:2025nev} at the Paul Scherrer Institute (PSI), beam injection and storage play a critical role due to the need for high muon decay statistics in these studies. In the Fermilab and PSI experiments, the challenge arises from the mismatch between the beam's phase space and the storage ring's or solenoid's acceptance phase space, compounded by the very narrow superconducting beam injection channel~\cite{Kim:2017jfd, Froemming:2019gcn, muEDM:2024bri}. As a result, achieving optimal beam storage efficiency has proven to be a daunting task in these experiments. 

Recently, Bayesian Optimization (BO)~\cite{7352306} has emerged as a valuable algorithm within the accelerator and beam physics community. It effectively addresses complex optimization challenges under noise and resource constraints during accelerator operation and resource-intensive beam physics simulations~\cite{Roussel:2023yin}. This algorithm employs probabilistic surrogate models and an acquisition function to balance exploration and exploitation, minimizing the number of evaluations. BO has been successfully applied in storage ring facilities, such as the Karlsruhe Research Accelerator (KARA)~\cite{Xu:2022ygq} and the Synchrotron Light Source DELTA at TU Dortmund University~\cite{Schirmer:2023vdn}. The Cooler Synchrotron storage ring COSY at Forschungszentrum Julich used BO for the optimization of the Injection Beam Line (IBL) to increase the beam intensity inside the storage ring~\cite{Awal:2023lpt}. With the advent of Laser Plasma Accelerators (LPAs), BO has been successfully demonstrated in simulations for concurrently optimizing the localized properties of compact free electron lasers (FELs) driven by laser wakefield accelerators (LWFAs), maximizing energy extraction efficiency and ensuring high-quality electron beams with reduced energy spread and emittance~\cite{Zhong:2024,Jiang:2025}. Additionally, BO has found applications in other domains, such as maximizing the Linac Coherent Light Source (LCLS) x-ray free-electron laser (FEL) pulse energy by controlling groups of quadrupole magnets~\cite{Duris:2019xwc} and optimizing multiple objectives in the MeV-ultrafast electron diffraction experiment at SLAC~\cite{Ji:2024gfq}. 

In this paper, we explore the use of BO to maximize storage efficiency in the Muon Electric Dipole Moment (muEDM) experiment at the Paul Scherrer Institute (PSI), focusing on optimizing the beam injection and storage within the storage solenoid of the experiment. The muEDM experiment at PSI aims to search for the muon electric dipole moment (EDM) with an unprecedented sensitivity of $6 \times 10^{-23}\,e\cdot$cm, which is four orders of magnitude better than the current limit~\cite{Muong-2:2008ebm}. Detecting a muon EDM larger than the Standard Model (SM) predictions~\cite{Cabibbo:1963yz, Kobayashi:1973fv, Pospelov:2013sca, Yamaguchi:2020eub} would provide an unambiguous hint of physics beyond the SM. Since the EDM violates time-reversal (T) symmetry, it also violates Charge-Parity (CP) symmetry, given that CPT symmetry is conserved. Therefore, the EDM can reveal new sources of CP violation, potentially shedding light on the matter-antimatter asymmetry observed in our universe~\cite{Sakharov:1967dj, WMAP:2012nax}.

\begin{figure}[htbp]
\includegraphics[width=0.85\linewidth]{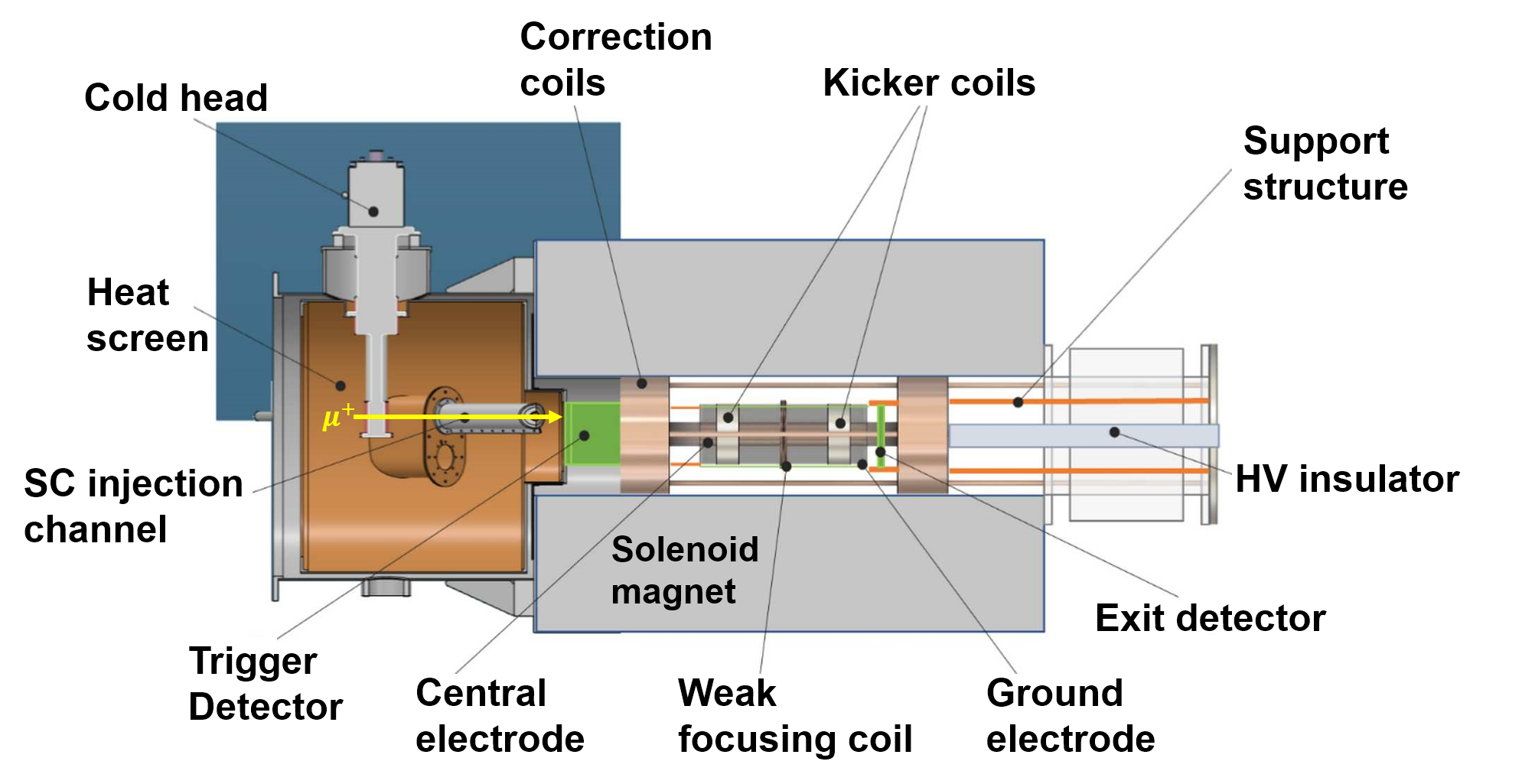}
\caption{A schematic view of the muEDM experiment at PSI for Phase I. It encompasses all components integral to the study presented herein, including the superconducting injection channel, the central electrode, kicker coils, and the weak focusing coil.}
\label{fig:muedm}
\end{figure}

The muEDM experiment consists of two phases: Phase I serves as a precursor experiment for the frozen-spin technique, aiming for a sensitivity goal of $4 \times 10^{-21}\,e\cdot$cm; and Phase II focuses on the muon EDM search with improved sensitivity ($6 \times 10^{-23}\,e\cdot$cm). In Phase I of the muEDM experiment~\cite{PSImuEDM:2023dsd}, the 28\,MeV/c surface muon beam at the $\pi$E1 beamline of PSI will be injected into a phase space compressor (PSC) solenoid with an inner diameter of 0.2\,m and a length of 1\,m, through a collimation tube with superconducting shield~\cite{muEDM:2023mtc} that are 15\,mm in diameter and 800\,mm long. The tube selects the appropriate phase space and shields the solenoid's fringe magnetic field. Inside the solenoid, five other coils generate a magnetic field to store the muons in the central region; these include a pair of correction coils, a weakly focusing coil, and a pulse coil~\cite{muEDM:2024bri}. When the muons exit the collimation tube and enter the 3-T magnetic field, they pass through a muon trigger detector~\cite{Hu:2025egg, Wong:2024vmo, Hu:2025tgk, Papa:2025tcu}, generating a signal that triggers the pulsed magnetic field, converting the longitudinal momentum of the muons into transverse momentum, which allows the muons to be stored in the weakly focused field. This beam injection scheme is motivated by the 3D spiral beam injection strategy~\cite{Matsushita:2023zho} of the J-PARC Muon $g-2$/EDM experiment~\cite{Abe:2019thb}. During storage, muons will circulate at a radius of $r = 31$\,mm with a cyclotron period of about 2.5\,ns until the muon decays into positrons and neutrinos. A radial electric field of 3\,kV/cm is applied by concentric cylindrical electrodes surrounding the muon orbit at $r = 40$\,mm (ground) and $r = 20$\,mm (high voltage) to satisfy the frozen-spin condition for measuring the muon EDM~\cite{muEDM:2024bri}. A schematic view of the experiment is shown in Fig.~\ref{fig:muedm}.

This paper is organized as follows: In Sec.~\ref{sec:injectparamstudy}, we provide an overview of the beam injection and storage process in the muEDM experiment and details about the simulation toolkit used in the optimization process. Section~\ref{sec:beaminjectionopt} presents optimization studies focused on identifying injection parameters that optimize the number of stored muons, including a comprehensive analysis of the BO technique. In Sec.~\ref{sec:result}, the results are discussed, and the implications of our findings are elaborated upon. Finally, this paper is summarized in Sec.~\ref{sec:summary}.

\section{Beam Injection and Storage Simulation}\label{sec:injectparamstudy}

\subsection{Beam Injection Phase Space at PSI’s $\pi$E1 Beamline}

\begin{figure}[htbp]
\centering
\includegraphics[width=0.485\linewidth]{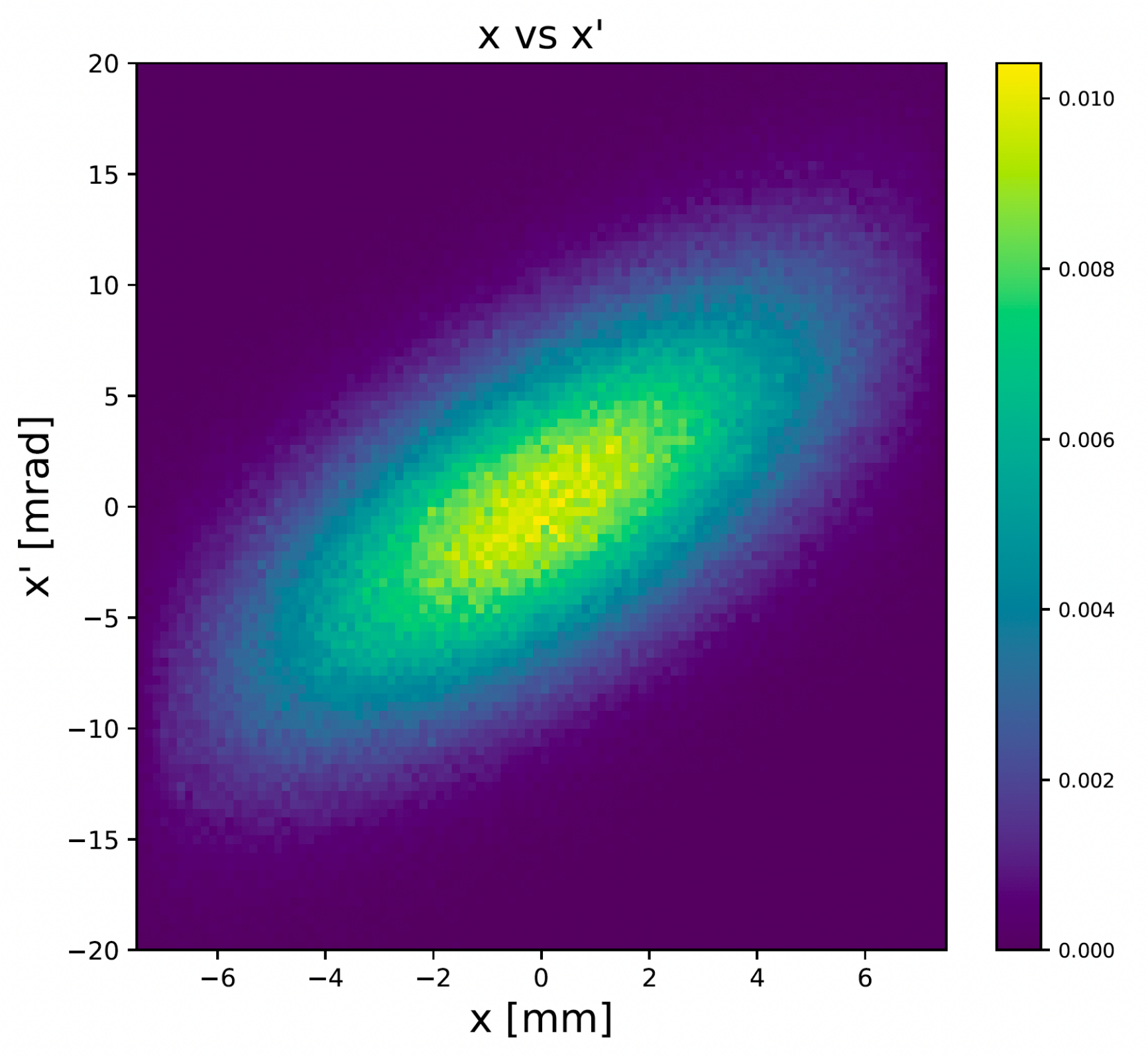}
\includegraphics[width=0.485\linewidth]{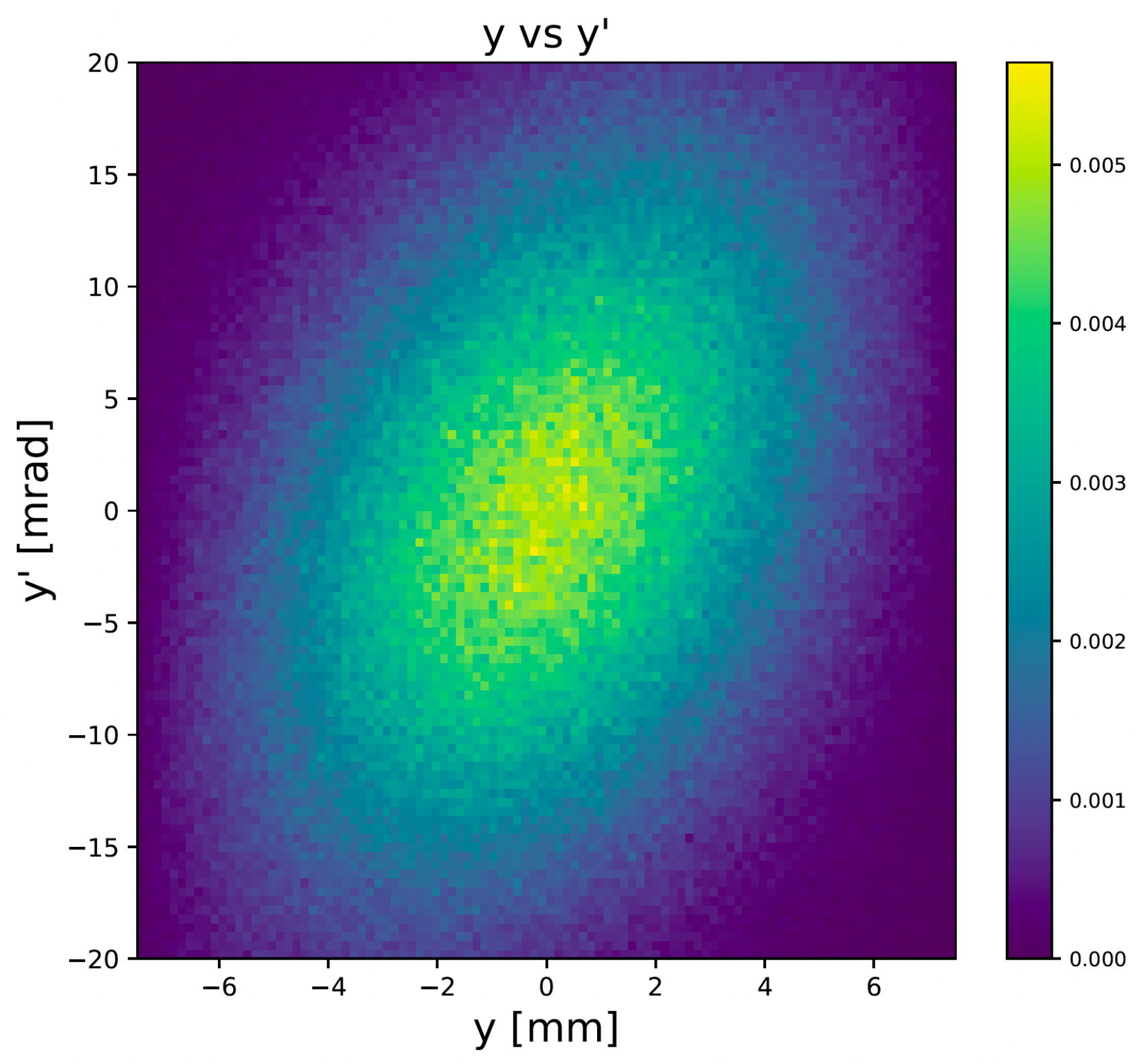}
\caption{Monte-Carlo generated $x$ (Left) and $y$ (Right) phase space after transmission through the injection tube with the total transmission of 3\%, with $x$ and $y$ emittance of 14.8\,mrad, and 32.0\,mrad respectively.}
\label{fig:bslDistInput}
\end{figure}

The $\pi$E1 beamline at PSI provides high-intensity pion and muon beams with momenta ranging from 10\,MeV/c to 500\,MeV/c and a momentum resolution better than 0.8\%. The beam is extracted from a carbon target, passes through dipoles and a Wien filter to select a muon beam with low contamination, and is tuned for the desired spin orientation. Beam transport to the experimental area uses quadrupole triplets for focusing and steering. Based on the measured transverse phase space of the $\pi$E1 beam~\cite{Papa:2015lda}, the transmission efficiency is manually optimized through simulation using injection tubes with different diameters $d_{\rm{inj}}$, indicating that $d_{\rm{inj}}$ = 15\,mm with a length of 800\,mm strikes a good balance in the selection and transmission of phase space. The beam profile after passing through the injection tube is shown in Fig.~\ref{fig:bslDistInput}.

\subsection{Simulation Framework and Modeling Tools}

The simulated phase space at the end of the injection tube, as depicted in Fig.~\ref{fig:bslDistInput}, was utilized to generate $5 \times 10^{6}\,\mu^{+}/s$ events for the subsequent simulation phase. G4Beamline~\cite{Roberts:2007nte} serves as a platform for prototyping the beam injection simulation, wherein the muon is introduced via an off-axis injection scheme into the PSC solenoid bore, which possesses dimensions of $x=200$\,mm in diameter and $z=1000$\,mm in length. The employed 3-T magnetic field is modeled by fitting empirical data to a calculated field, then adjusting solenoid and split coil pair parameters within an ANSYS simulation~\cite{Alawadhi:2015}.

\begin{figure}[htbp]
    \centering
    \includegraphics[width=0.85\linewidth]{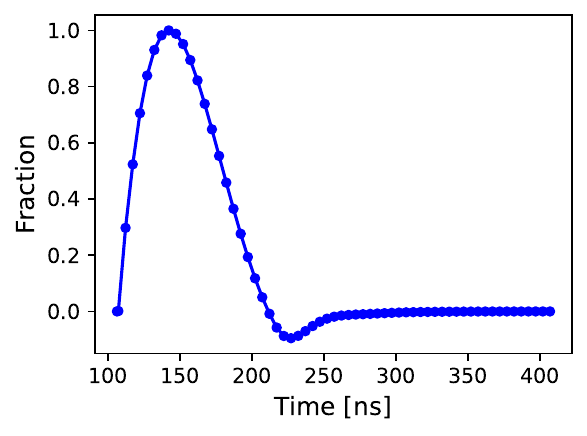}
    \caption{Kicker pulse current profiles simulated using the LTspice circuit model. The simulated time range is from $t=0$ to $t=500$\,ns, whereas $t=0$ signifies the beam entrance into the storage solenoid.}
    \label{fig:kicker}
\end{figure}

A pair of correction coils, with inner and outer radii measuring 90\,mm and 99.9\,mm respectively, along with a length of 90\,mm, is powered by a current of 2.5\,A/mm$^{2}$. These coils are simulated at position $z=\pm 250$\,mm to enhance the acceptance of the injection phase space between the exit of the injection tube and the storage range. Additionally, a magnetic coil with inner and outer radii of 50\,mm and 60\,mm, and a length of 10\,mm, is positioned 100\,mm apart with anti-parallel currents. This coil is simulated at the center to create the pulsed magnetic kicker field. The configuration of the magnetic pulsed kicker, as depicted in Fig.~\ref{fig:kicker}, is modeled based on a simulation derived from the circuit design tool LTspice~\footnote{https://www.analog.com/en/design-center/design-tools-and-calculators/ltspice-simulator.html}.

The weakly focusing coil, with an inner radius of 50\,mm, an outer radius of 60\,mm, and a length of 10\,mm, is simulated at the center to provide longitudinal confinement for the muon with a current of 1.5\,A/mm$^{2}$. The coaxial electrodes, which are 120\,mm long and made of carbon and copper, consist of a high-voltage (HV) and a ground electrode. These electrodes create the electric field necessary for maintaining the frozen spin condition. The magnetic fields generated by both the correction and the weakly focusing coil are modeled using G4Beamline, while the electric field is modeled using ANSYS.

The simulation models the muon injection process into a storage region, considering various injection parameters that influence beam dynamics and efficiency. The injection geometry is defined by the injection angle, $\theta$, and transverse angle, $\phi$, measured in degrees, determining the muon’s entry trajectory. The injection angle, $\theta$, is the angle formed by the injection tube with respect to the Z-axis, while the transverse angle, $\phi$, is the angle measured relative to the Y-axis, located at the rear side of the PSC solenoid. The injection radius, $R_{\rm{inj}}$, and the longitudinal injection coordinate, $Z$, both measured in millimeters, specify the spatial entry point. The injection radius, $R_{\rm{inj}}$, is the radial distance from the rear center to the desired injection circumference. Furthermore, the weakly-focusing coil current, $A_{\text{weak}}$, expressed in Amperes per millimeter (A/mm), is vital for shaping the field to confine the muon beam. The kicker field strength, BPI, and the pulsed kicker time offset (KPT), measured in nanoseconds (ns), is relative to the shape of the kicker pulse Fig.~\ref{fig:kicker}, regulate beam steering and timing, ensuring optimal injection and subsequent storage conditions. Collectively, these beam injection and storage parameters are summarized in Tab.~\ref{tab:injParam}.

\begin{table}[htbp]
\caption{List of parameters used in the baseline simulation optimization study.}
\label{tab:injParam}
\begin{tabular}{cc}
\hline
Parameter symbol & Description           \\
\hline
$R_{\rm{inj}}$  & Injection radius (mm)       \\
Z           & Longitudinal injection coordinate (mm) \\
$\theta$    & Injection angle (degree)    \\
$\phi$      & Transverse angle (degrees)  \\
$A_{weak}\times 100$  & Weak current $\times$ 100 (A/mm)         \\
BPI         & Strength of pulsed kicker (arb. units)   \\
KPT         & Time offset of pulsed kicker (ns)     \\
\hline
\end{tabular}
\end{table}

\begin{figure}[htbp]
    \centering
    \includegraphics[width=0.95\linewidth]{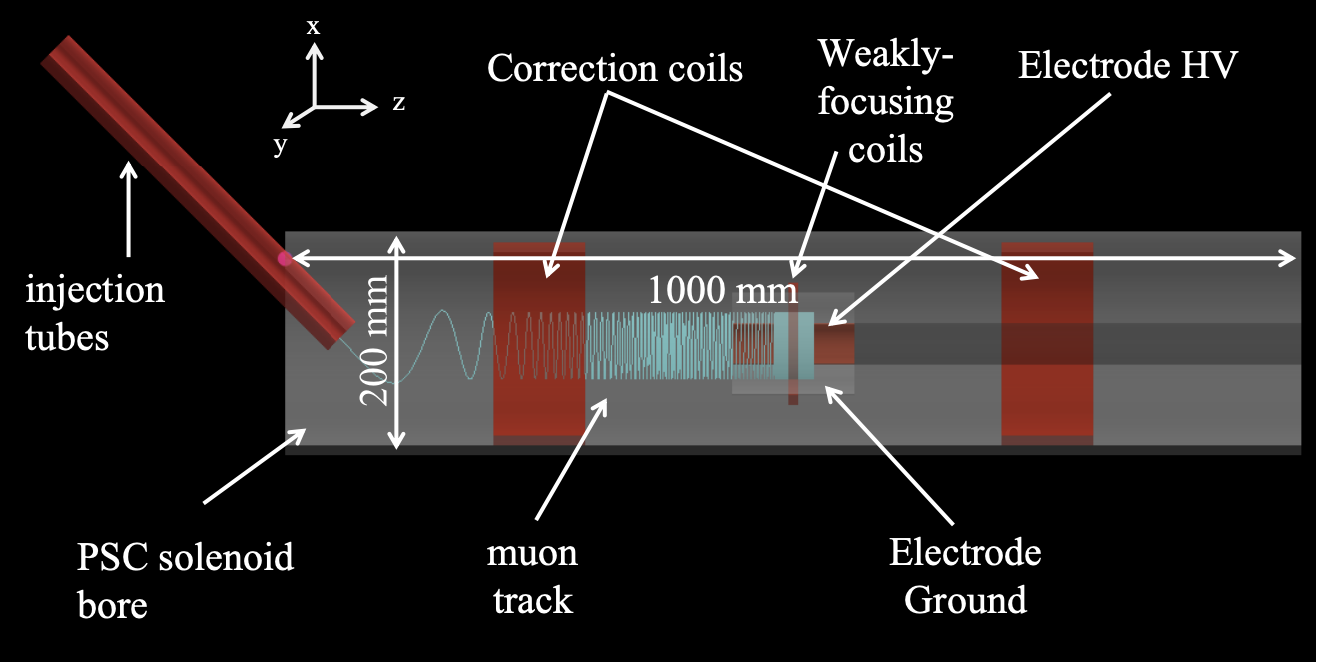}
    \caption{Visualization of a beam injection and storage event into the PSC magnet in the G4Beamline simulation. The image depicts a ``pencil'' muon beam trajectory inside the solenoid bore. Key components are labeled, including the injection tube, correction coils, weakly focusing coils, and high-voltage electrodes, which play crucial roles in guiding and storing the muon beam within the 1000 mm-long solenoid.}
    \label{fig:PSC_solenoid}
\end{figure}

For a specified set of injection parameter inputs, the stored muon efficiency, $\epsilon$, is given by the equation: 

\begin{equation}
\epsilon = \frac{N_{\rm{stored}}}{N_{\rm{injected}}}
\end{equation}
where the $N_{\rm{stored}}$ is the number of stored muons retained in the central region given by $|z|<40\,\text{mm}$. On top of that, these muons shall remain in this region for more than 300\,ns, given that the total elapsed time from their exit from the injection tube to their arrival in the central region is less than 200\,ns. Additionally, $N_{\rm{injected}}$ represents the total number of muons injected. In other words, the number of muons passed through the injection tube. A typical stored muon event, as simulated within the G4beamline framework, is illustrated in Fig.~\ref{fig:PSC_solenoid}.

\section{Optimization Methodology}\label{sec:beaminjectionopt}
\subsection{Initial Optimization with Polynomial Chaos Expansion}\label{sec:PCE}

Since the BO routine is very sensitive to the initial parameter choices and bounds, we employed a surrogate model based on Polynomial Chaos Expansion (PCE) to expedite our process before utilizing BO for optimal beam injection and storage parameters computation.

PCE is a spectral expansion method that expresses the response of a system as a series of orthogonal polynomials of input parameters, providing an efficient surrogate modeling approach for complex physical systems. It has been widely used in uncertainty quantification and parametric sensitivity analysis, serving as an alternative to direct numerical simulations, which can be computationally expensive. The surrogate model PCE encodes the responses to input variables, where their distributions are related through an expansion coefficient. This coefficient is calculated using non-intrusive methods, estimated by regression techniques based on the difference between the output predicted by the model and the true response provided by the simulation results. The regression-based estimation of coefficients depends on the number of samples and the integration points.

An 8-dimensional PCE model~\cite{muEDM:2023xkb} based on the injection parameters from Tab.~\ref{tab:injParam} incorporating the kicker width as an additional variable. Additionally, an initial distribution of $10^{6}$ muons was used for this iteration. Assuming a binomial distribution for successful injection would translate to a variance of $\approx 2.0$\% for an injection efficiency of $\approx 0.3$\%. Thus, the accuracy in this case was limited by the small number of training samples and the number of muons in the initial distribution used for injection. A 3rd or 4th-degree polynomial expansion was trained using 1,600 samples generated from G4Beamline simulations. The mean squared error (MSE) for the 3rd or 4th degree PCE model was in the range of 10$^{-6}$ to 10$^{-7}$, for the number of training samples. Utilizing the ChaosPy Python toolbox~\cite{Feinberg:2015}, the generated PCE expansion is fitted with the trained samples using the least squares regression method~\cite{muEDM:2023xkb}. The optimal value of storage efficiency, as determined, is 0.324\%. Although this approach has the potential for further enhancement by including additional samples in the training phase, specifically, exceeding 3000 samples for the fourth degree polynomial and over 5000 for the sixth order polynomial, we have opted to conclude our efforts at this juncture. This decision is predicated on the consideration that further expansion would incur excessive computational costs. In contrast, our primary objective was identifying reasonable initial parameters and their respective boundaries for the BO routine. 

\subsection{Bayesian optimization with Gaussian process}

Bayesian Optimization (BO) is a powerful strategy for global optimization, particularly effective for evaluating expensive, noisy and time-intensive objective function~\cite{Brochu:2010lgg}. It uses a probabilistic surrogate model to navigate the search space by employing a sampling strategy that balances probing uncertain areas with focusing on regions likely to yield optimal results. In each iteration, BO selects a new evaluation point based on this criterion, refines the surrogate model with recent observations, and repeats the cycle. This iterative process incrementally improves the model and guides the optimization toward the global optimum. The BO algorithm used in this study follows the structure outlined in Algorithm~\ref{tab:algo}~\cite{Brochu:2010lgg}.

\begin{algorithm}
\caption{Bayesian optimization algorithm}\label{tab:algo}
\KwData{$t \geq 0$}
\KwResult{$D_{t} = [X,Y]_{t}$}
$Y \gets f_{obj}(\bf{x_{n}})$\;
$\bf{X} \gets \bf{x_{0}}$\;
$N \gets 1$\;
\While{$N < t$}{
    $\bf{X} \gets \text{argmax } \alpha(\bf{x_{t-1}}|D_{t-1})$\; 
    $Y \gets f_{obj}(\bf{X})$\;
    $N \gets N + 1$\;
}
\end{algorithm}

A Gaussian Process (GP) is often chosen as the surrogate, as it provides a non-parametric, probabilistic model of the objective function and is commonly used together with an acquisition function to guide the search for the next evaluation point~\cite{Rasmussen:2005,Xu:2022ygq}. The GP defines a distribution over functions such that any finite collection of function values has a joint Gaussian distribution $\mathcal{N}(\mu,\sigma)$:
\begin{equation}
\mathbf{f(x)} = \left[ f(x_{1}) , f(x_{2}), ..., f(x_{N}) \right]^{T}
\sim \mathcal{N} ( \mu(x) , k( \bf{x} , \bf{x^{\prime}} ) )
\label{eq:gp}
\end{equation}
characterized by the mean function $\mu(x)$, representing the expected value of the function at input $\bf{x}$, and a covariance function, or kernel $k(\bf{x},\bf{x}')$. The choice of the kernel encodes the objective function's assumptions. In this case, the Radial Basis Function $k_{RBF}( \bf{x}, \bf{x}'; \ell)$ is chosen, valued for its smoothness and adaptability~\cite{MAJDISOVA2017728}. This kernel captures correlations based on Euclidean distances, allowing the model to accommodate variations in function behavior across different dimensions. To account for observational noise in the model, the kernel is modified by adding a Gaussian noise term to its diagonal. The resulting expression for the kernel becomes:
\begin{equation}
k(\bf{x}, \bf{x}') = \sigma^2 k_{RBF}( \textbf{x}, \textbf{x}'; \ell) + \sigma^2_{\text{noise}} \delta_{ij}
\label{eq:fitfunction}
\end{equation}
In this formulation, $\sigma^2$ represents the signal variance, $\sigma^2_{\text{noise}}$ captures the variance of the additive Gaussian noise, and $\delta_{ij}$ is the Kronecker delta function ensuring that noise affects only the diagonal elements. 

The length scale $\ell$ prior distribution reflects initial assumptions about the function's behavior. As new data is incorporated, these assumptions are iteratively refined. The output distribution of the objective function is progressively updated through Bayesian inference. Meanwhile, the objective function $f(\textbf{x})$ is modeled as GP by updating the data used and performing the Bayesian regression:
\begin{equation}
    p(\mathbf{\mathbf{f(x)}}|\mathcal{D}) = \mathcal{GP} \left( \mathbf{f} ; \mu(\mathbf{x})_{\mathbf{f}|D} , k(\mathbf{x},\mathbf{x^{\prime}})_{\mathbf{f}|D} \right)
\end{equation}
where $D$ is the dataset,
\begin{equation}
    D = \{(x_{1},y_{1}),(x_{2},y_{2}), ..., (x_{N},y_{N})\}
\end{equation}
The acquisition function determines the next evaluation point in the parameter space. This function is designed to balance exploration, sampling regions where uncertainty is high, and exploitation, sampling regions where the predicted value is optimal. Among the commonly used acquisition functions are the Upper Confidence Bound (UCB) and Expected Improvement (EI), each offering different strategies to balance exploration and exploitation~\cite{7352306}. In this work, we adopt the UCB approach due to its simplicity and tunable trade-off between exploration and exploitation, which is well-suited for our problem setup. There are given by:
\begin{equation}
    \alpha_{UCB}(x;\kappa) = \mu (x)+\kappa \sigma(x)
\end{equation}
where $\mu(x)$ is the predicted mean and $\sigma(x)$ is the predictive uncertainty of the GP function. High $\kappa$ values emphasize the uncertainty term, promoting exploration; conversely, small $\sigma(x)$ values reduce the impact of $\sigma(x)$, favoring exploitation by sampling near regions with higher posterior mean values given by the observed peaks. This adaptive strategy enables BO to efficiently navigate complex, high-dimensional parameter spaces with minimal evaluations. The $\kappa$ can also
increase along with the evaluation steps to ensure that BO converges to the global optimum~\cite{Brochu:2010lgg,6138914}.

\section{Optimization Results}\label{sec:result}

\begin{figure}[htbp]
    \centering
    \includegraphics[width=0.75\linewidth]{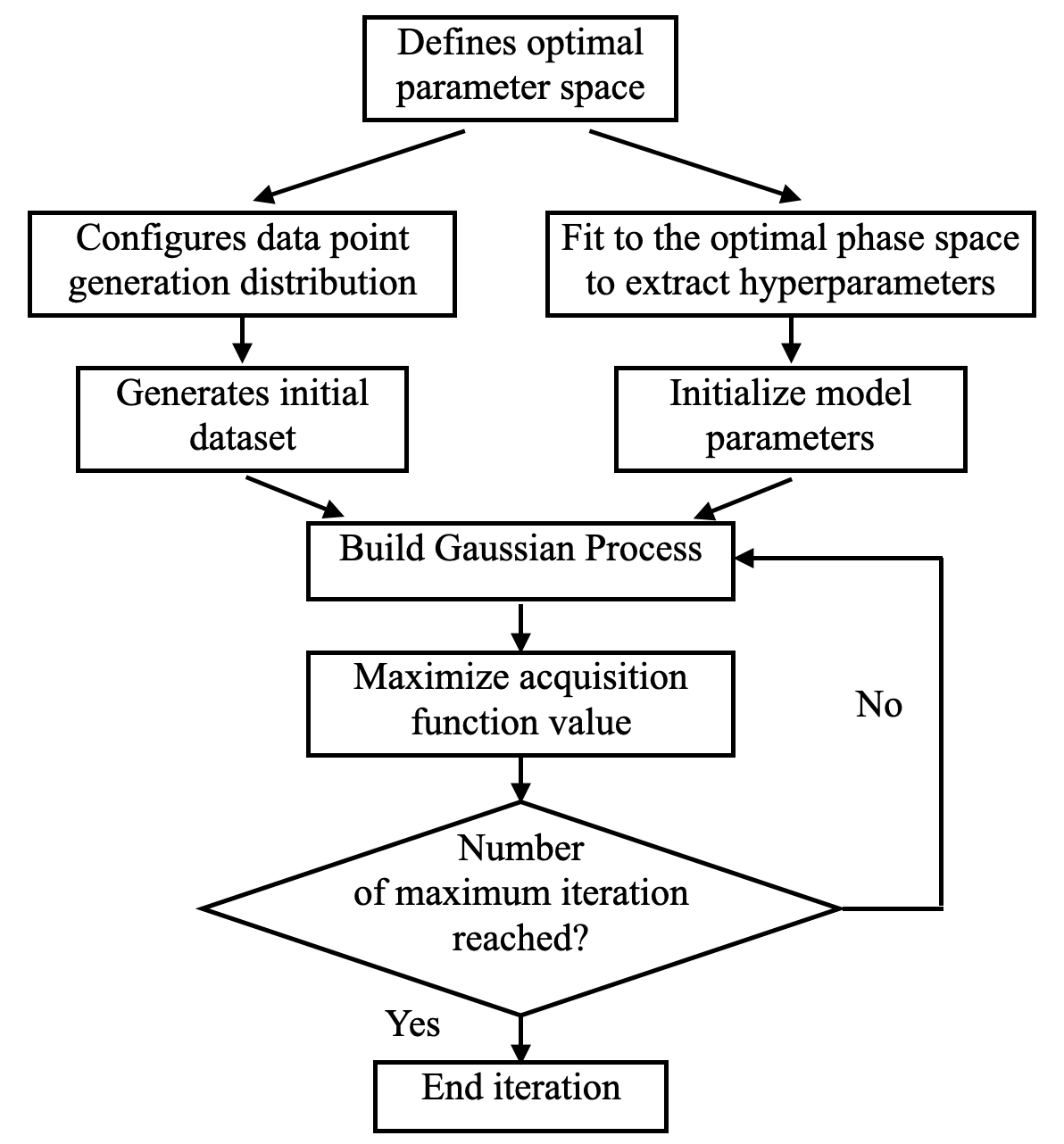}
    \caption{The workflow of BO implemented in the beam storage and injection study.}
    \label{fig:bo}
\end{figure}

The optimization of beam injection and storage efficiency follows the workflow shown in Fig.~\ref{fig:bo}. The process begins by defining an optimal parameter space, which is used both to configure the data point generation distribution and to extract model hyperparameters. Following this, the GP model is constructed and iteratively refined by maximizing an acquisition function within the defined parameter space. At each step, the acquisition function proposes new sampling points, which are evaluated and incorporated to update the surrogate. This loop continues until a predefined maximum number of iterations is reached, thus concluding the optimization process.

\begin{figure}[htbp]
    \centering
    \includegraphics[width=0.85\linewidth]{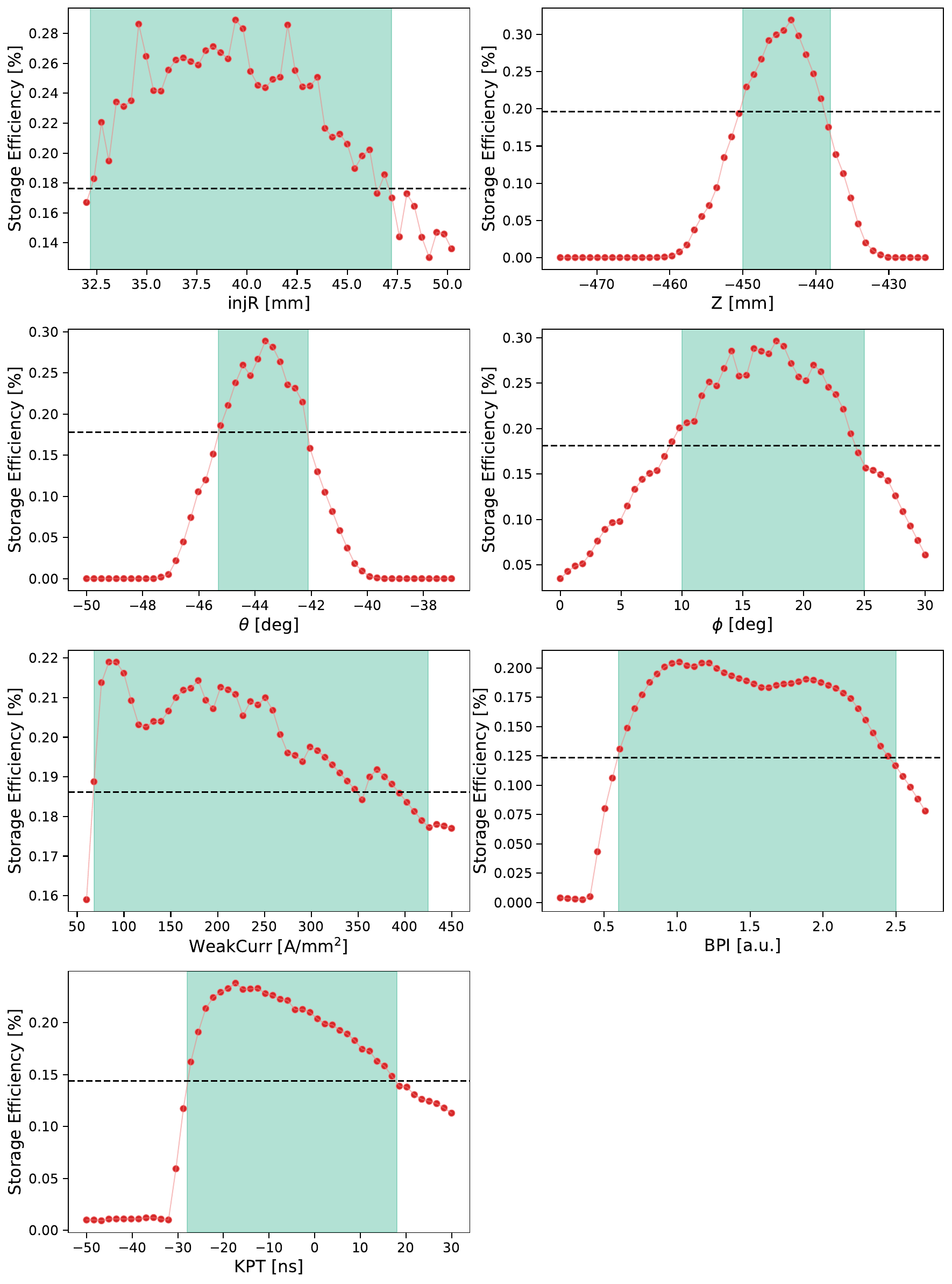}
    \caption{Parameter scan for each injection parameter, with other parameters kept constant during the scan to obtain the sensitive region. The green highlighted region represents the range given by 60\% of the maximum efficiency, in which an optimal range is obtained. The optimal range for weak current parameter is given by 85\% coverage.}
    \label{fig:scan_param}
\end{figure}

The storage efficiency of injected muons is influenced by the initial parameters (Tab.~\ref{tab:injParam}). Each parameter is independently scanned using G4Beamline simulations, with the remaining parameters fixed at their baseline value (PCE-optimized set of parameter, shown in Tab.~\ref{tab:comparion_PCE_BO}). For each parameter, we define the optimal range as where efficiency exceeds 60\% of the maximum observed value, applying a stricter 85\% threshold for parameters with flatter response profiles (Figure~\ref{fig:scan_param}).

\begin{figure}[htbp]
    \centering
    \includegraphics[width=0.85\linewidth]{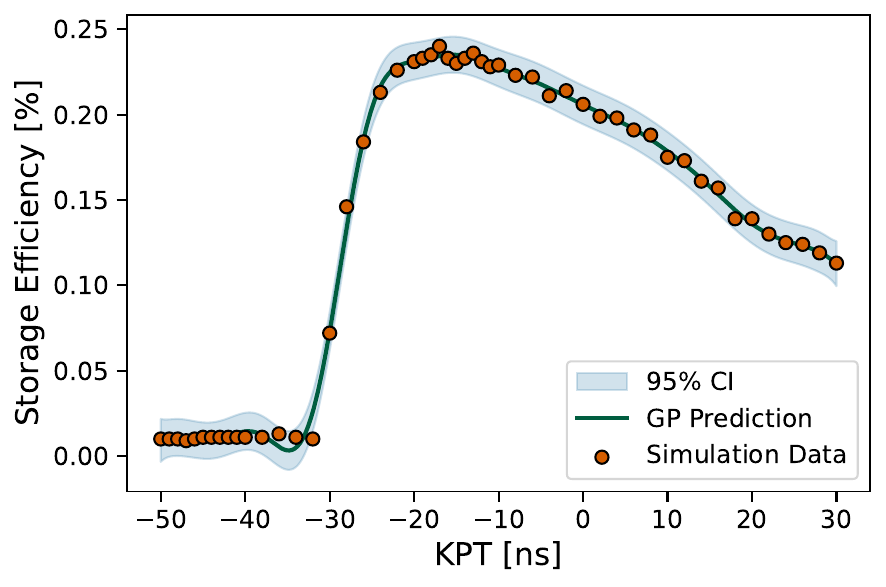}
    \caption{An example of a fit to the KPT scan profile, in which the GP hyperparameter ${\ell, \sigma^{2}, \sigma_{noise}^{2}}$ are extracted via the log marginal likelihood method. The solid line green line shows the posterior mean prediction, with the shaded region indicates the 95\% confidence level.}
    \label{fig:GP_KPT}
\end{figure}

Traditional Monte Carlo methods rely on random sampling, which can lead to clustering and gaps in parameter coverage, this is particularly problematic in our seven-dimensional injection efficiency study where sparse sampling risks missing critical optima. By contrast, quasi-random low-discrepancy sequences systematically fill the parameter space with maximally uniform point distributions~\cite{Cheng:2013}. For our initial dataset $D_{0}$, we generate ten optimally spaced points via Sobol sequence~\cite{SOBOL196786} based sampling, the storage efficiency for each sampled point is then evaluated through simulation. 
Additionally, we fit each parameter scan profile to Eq.~\ref{eq:fitfunction} using log marginal likelihood maximization to extract the GP hyperparameters, shown in Tab.~\ref{tab:HPfit}, an example of the fit is shown in Fig.~\ref{fig:GP_KPT}.

\begin{table}[htbp]
\caption{Hyperparameters length scale normalized to the optimal search space ($\ell$), signal variance ($\sigma^{2}$), and noise ($\sigma_{noise}$), extracted for each parameter.}
\label{tab:HPfit}
\begin{tabular}{lcccc}
\toprule
Parameter & $\ell$ & $\sigma$ & $\sigma_{noise}^{2}$  \\
\hline
$R_{\rm{inj}}$ (mm)     & 0.753 & 0.174 & 2.98e-4 \\
Z (mm)             & 0.685 & 0.194 & 1.55e-4 \\
$\theta$ (degree)  & 0.737 & 0.171 & 1.75e-4 \\
$\phi$ (degree)    & 0.807 & 0.183 & 3.13e-4 \\
$A_{weak}\times 100$ ($\rm{A/mm^{2}}$)  & 0.141 & 0.163 & 6.33e-6 \\
BPI (arb. unit)    & 0.200 & 0.176 & 6.33e-6  \\
KPT (ns)           & 0.143 & 0.172 & 7.96e-6 \\
\hline
\end{tabular}
\end{table}

The $A_{weak}$, KPT, and BPI exhibited the smallest length scale in the GP model, indicating that storage efficiency is highly sensitive to this parameter variations. This aligns with expectations, as the KPT and BPI characterize the pulsed kicker 
employed to trap muons in the central region of the solenoid; and $A_{weak}$ is important used in confining the muon. In contrast, geometric parameters such as $R_{\rm{inj}}$, Z, $\theta$ and $\phi$ have larger length scales, suggesting a reduced sensitivity to minor changes in their values. The signal variances for all parameters are relatively similar, implying comparable levels of functional variation among them. The noise levels remain low across all parameters, consistent with well-controlled experimental or simulation conditions setup.

\begin{figure}[htbp]
    \centering
    \includegraphics[width=0.85\linewidth]{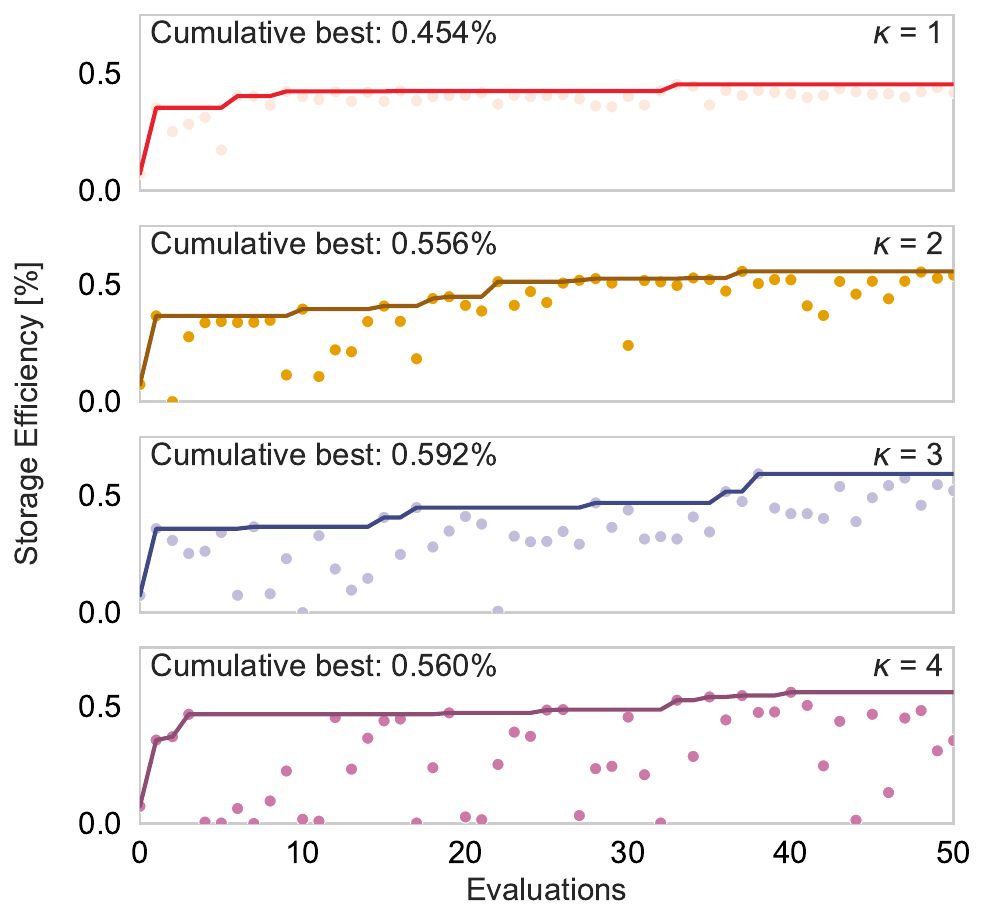}
    \caption{The muon storage efficiency for different $\kappa$ values, with the cumulative best results for 50 iterations.}
    \label{fig:kappa_study}
\end{figure}

The GP surrogate model is constructed using the $D_{0}$ and the fitted hyperparameters. The mean function in the GP is commonly set to $\mu(x)$ = 0, a standard assumption when the objective function is unknown or lacks a defined prior~\cite{Xu:2022ygq}, ensuring a non-informative yet flexible starting point for the surrogate model. To improve the performance of BO, different values of the confidence parameter $\kappa$ were explored, as shown in Fig.~\ref{fig:kappa_study}. The increase in $\kappa$ from 1 to 3 has seen the optimizer to explore wider into other region of the phase space, as evident by the sparsely scattered storage efficiency points with respect to the cumulative best results as $\kappa$ increases. However, the optimizer performance started to degrade at $\kappa$ = 4. Consequently, a confidence parameter of $\kappa$ = 3, corresponding to the 99.7\% confidence interval of a Gaussian distribution~\cite{Xu:2022ygq}, was chosen for the final configuration. On the other hand, the number of iterations was increased from 50 to 100 to evaluate potential improvements. However, the improvement is marginal, at this point, only iteration 50 is chosen to conserve computation time.

The BO algorithm was implemented using the GPy software package~\footnote{University of Sheffield Machine Learning group, A gaussian process framework in python, https://github.com/SheffieldML/GPy} to construct the GP model, and the SciPy library~\cite{Virtanen:2019joe} was used to optimize the acquisition function. Each iteration of BO involved 100,000 injection events, and the process was run for 50 iterations, requiring approximately 120 hours in total. Notably, most of the computation time per iteration was spent executing the simulation.

\begin{figure}[htbp]
    \centering
    \includegraphics[width=0.95\linewidth]{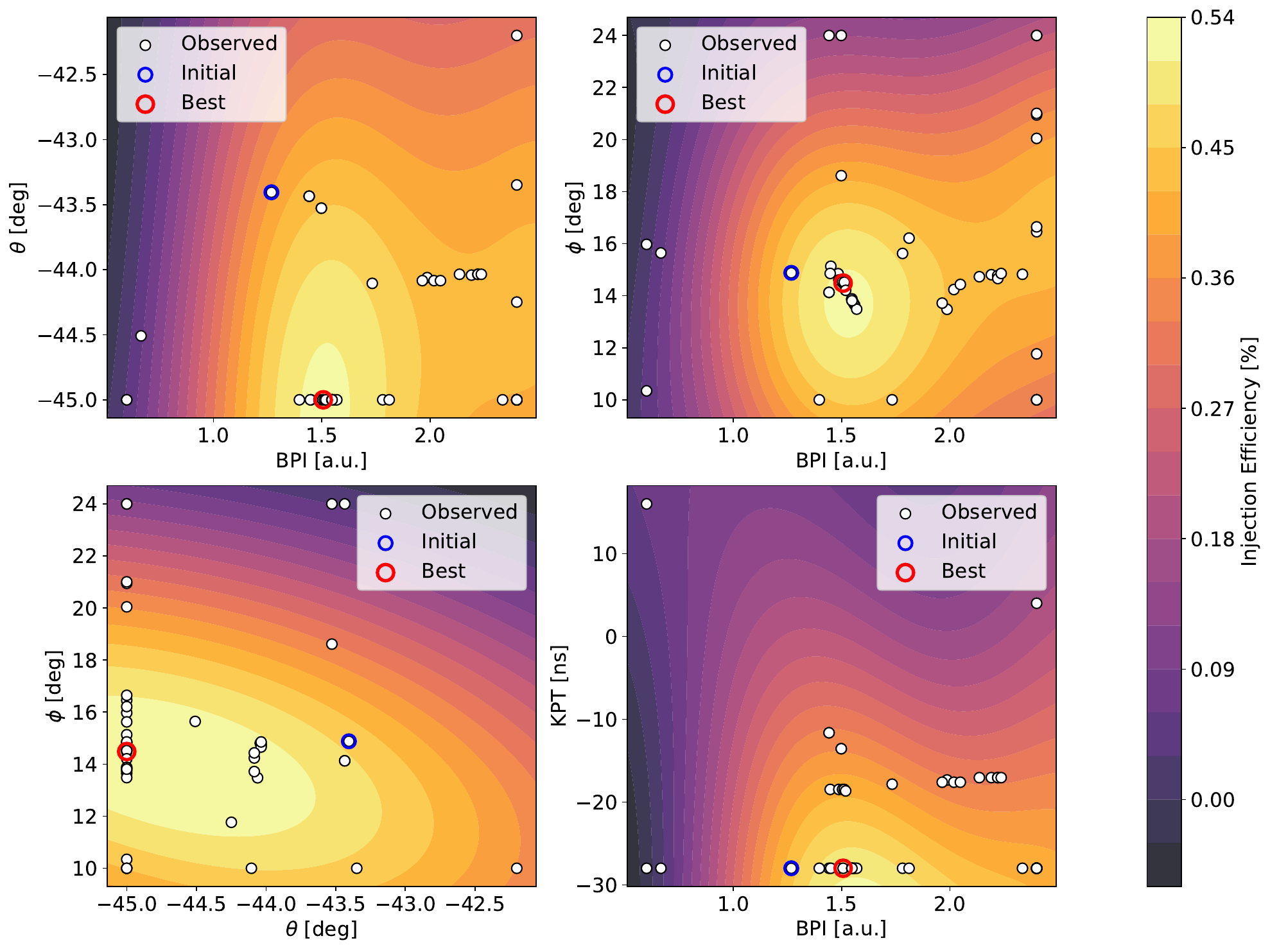}
    \caption{The performance of BO in optimizing the sensitive parameters as it attempting to ascend to the maximal GP posterior mean. The blue circle point indicates the initial point, white circle points are the observed point throughout iteration, and the red circle point is the final point.}
    \label{fig:Contour_2D}
\end{figure}

\begin{figure}[htbp]
    \centering
    \includegraphics[width=0.7\linewidth]{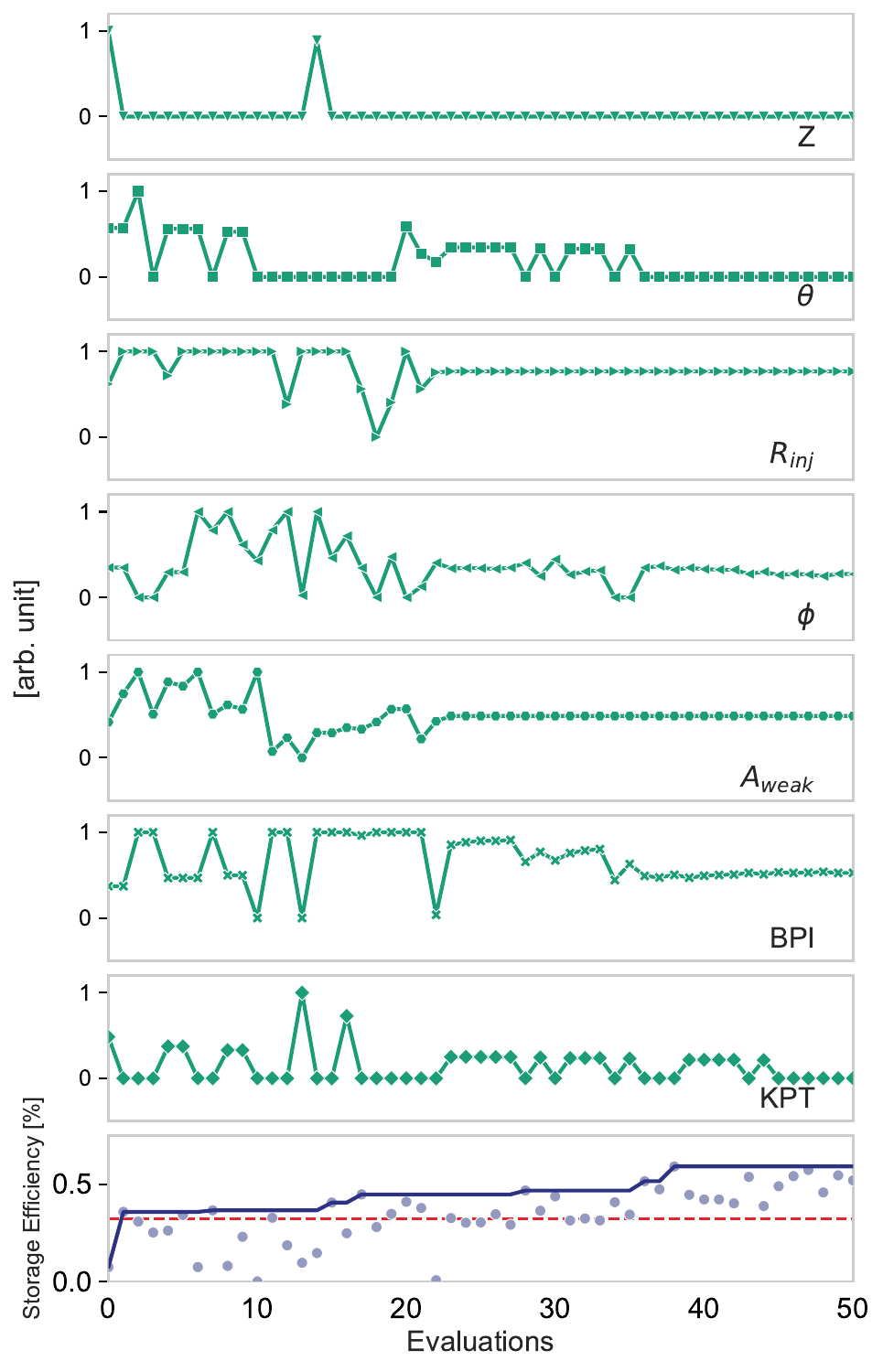}
    \caption{The evolution of the injection parameters and the resulting objective function are illustrated across the optimization steps, with the vertical limit representing the total allowed optimization range (normalized). The bottommost plot displays the muon storage efficiency (light purple markers) alongside the cumulative best results (dark purple line). The red dotted line indicates the muon storage efficiency obtained from the PCE optimization algorithm.}
    \label{fig:result}
\end{figure}

The optimization results, along with the parameter evolution across iterations, are shown in Fig.~\ref{fig:result}. Most parameters exhibit fluctuations early in the optimization process, reflecting exploration during the Bayesian Optimization. As the iterations progress, several parameters, namely $Z$, $R_{\rm{inj}}$, and $A_{\rm{weak}}$ tend to stabilize, while the others continue to vary, suggesting ongoing exploration of their effects on storage efficiency. On the other hand, Fig.~\ref{fig:Contour_2D} shows the example of sensitive parameters optimal point being optimized in the efficiency landscape predicted by the GP surrogate model. Through iteration, the observed optimized parameters are converging to the brighter region, signifying maximal efficiency region. Also, it reveal how the optimizer balances exploitation of promising configurations (clustered points) with exploration of uncertain regions (dispersed points). These observation, together with the increase in storage efficiency throughout the iterations confirms the optimization process's success in achieving its goal. 

A comparison between the optimized results obtained using the PCE method (Sec.~\ref{sec:PCE}) and those from BO is presented in Tab.~\ref{tab:comparion_PCE_BO}. The BO optimization improved the storage efficiency nearly twofold, from 0.324\% (PCE) to 0.596\%. The results were cross-validated using the \texttt{musrSim} simulation package~\cite{Sedlak:2012}, confirming the consistency of the optimized parameter set and the corresponding storage efficiency.

\begin{table}[htbp]
\caption{Optimized injection parameters for the PCE and BO method.}
\label{tab:comparion_PCE_BO}
\begin{tabular}{lcc}
\hline
Parameter                               & PCE      & BO  \\
\hline
$R_{\rm{inj}}$ (mm)                     & 45.56   & 43.57  \\
Z (mm)                                  & -443.84 & -450.00  \\
$\theta$ (degree)                       & -45.02  & -45.00  \\
$\phi$ (degree)                         & 9.24    & 14.49  \\
$A_{\rm{weak}}\times 100(\rm{A/mm^2})$  & 150.00      & 210.98  \\
BPI (arb. unit)                         & 1.00        & 1.507   \\
KPT (ns)                                & 0        & -28  \\ \hline
Storage efficiency (\%)                 & 0.324    & 0.592 \\
\hline
\end{tabular}
\end{table}

This improvement comes with specific trade-offs. The increase in current for the weakly focusing coil enhances beam-focusing capabilities, but it also leads to higher energy consumption. The rise in kicker field strength enhances the muons' stopping in the storage region, imposing additional demands on the kicker system, such as greater power requirements and potential wear over extended operational periods. Conversely, the -28\,ns time offset for the pulsed kicker effectively advances the timing of the kicker activation. This earlier activation helps synchronize the kicker's operation with the beam injection process, thereby improving muon capture efficiency. While this adjustment optimizes timing, it may necessitate more precise control and synchronization of the system to sustain consistent performance under varying conditions.

\section{Summary and Outlook}
\label{sec:summary}
This study demonstrates using Bayesian Optimization to improve beam injection and storage efficiency for the muEDM experiment at the Paul Scherrer Institute (PSI), which focuses on searching for the muon electric dipole moment. By optimizing a defined set of injection parameters through simulations, the BO framework effectively improves storage efficiency while minimizing the number of evaluations conducted. This work also shows the method's feasibility and establishes a foundation for optimizing beam injection and storage in both Phase I and Phase II of the muEDM experiment, with the current implementation serving as a baseline framework for future development. Although the present BO approach may impose higher demands on the experiment's subsystems involving currents, incorporating physical safety constraints can confine exploration to safe parameter regions, thereby addressing these challenges. 

Future improvements to the Bayesian Optimization (BO) process could focus on refining the selection of the scan range used for sampling, which directly impacts the balance between exploration and exploitation. Optimizing this range may enhance the efficiency and coverage of the parameter space. Furthermore, extending the current single-objective BO framework to a multi-objective optimization approach would enable simultaneous trade-offs among competing experimental goals; for instance, maximizing storage efficiency while minimizing complexity in correction coil geometry and current requirements.

\begin{acknowledgments}
The computations presented in this paper were performed on the Siyuan-1 cluster, supported by the Center for High Performance Computing and the Institute of Nuclear and Particle Physics (INPAC) Cluster at Shanghai Jiao Tong University. This work is supported by the National Natural Science Foundation of China (No.12050410233) and Shanghai Jiao Tong University Double First Class startup fund (WF220442608). Y.~Takeuchi acknowledges support from the K.~C.~Wong Educational Foundation.
\end{acknowledgments}

\bibliography{reference}

\end{document}